\documentclass[12pt]{iopart}

\usepackage{graphicx}
\begin{document}

\title[Compact Object Formation]{Compact Object Formation and 
the Supernova Explosion Engine}

\author{Chris L. Fryer}

\address{CCS-2, MS D409, 
Los Alamos National Laboratory
Los Alamos, NM, 87544}
\ead{fryer@lanl.gov}

\begin{abstract}

When a massive star ends its life, its core collapses, forming a neutron 
star or black hole and producing some of the most energetic explosions in 
the universe.  Core-collapse supernovae and long-duration gamma-ray bursts 
are the violent signatures of compact remnant formation.  As such, both 
fields are intertwined and, coupled with theory, observations 
of transients can help us better understand compact remnants just as 
neutron star and black hole observations can constrain the supernova 
and gamma-ray burst engine.  We review these ties in this paper.

\end{abstract}

%Uncomment for PACS numbers title message
%\pacs{00.00, 20.00, 42.10}
% Keywords required only for MST, PB, PMB, PM, JOA, JOB? 
%\vspace{2pc}
%\noindent{\it Keywords}: Article preparation, IOP journals
% Uncomment for Submitted to journal title message
%\submitto{\JPA}
% Comment out if separate title page not required
\maketitle

\section{The progenitors of Compact Remnants}

In molecular clouds, clumps of gas can become gravitationally bound
and contract.  Ultimately, the compressing gas heats up sufficiently 
to drive the fusion of hydrogen into helium and the birth of a star.  
For most of its life, hydrogen fusion powers the star (the so-called 
main sequence evolution).  Ultimately, the hydrogen is depleted in 
the core and the star expands into a giant phase as its core contracts.  
For massive stars, this phase represents just the first in a series 
of burning phases where the ashes of the first phase ignite as the 
core contracts:  core contraction continues until helium fusion 
occurs, a second core contraction until the products of helium 
fusion (carbon, oxygen) ignite, and so on until an iron core 
is built in the center of the star.  Silicon shell burning above 
the iron core continues to increase its mass until the thermal 
and electron degeneracy pressure in the core is no longer able 
to support its mass and the core collapses.

The collapse of these cores produce most of the stellar-massed neutron
stars (NSs) and black holes (BHs) in the universe\footnote{A subset of neutron
  stars are produced by the collapse of ONe cores that collapse before
  creating an iron core - the so-called electron capture supernovae.
  In a similar manner accreting ONe white dwarfs can collapse if the
  accrete enough material in an interacting binary.  The collapse of
  these systems is similar to the iron cores, and we will focus on the
  iron core collapse scenario in this review.}.  The energy released
in these core collapses produce supernovae (type Ib/c and II) and
long-duration gamma-ray bursts (GRBs).  The nature and mass of the remnants
is directly connected to the nature of the explosion.  In this paper,
we review the current models for the explosions from stellar implosion 
and the resultant role they play on the compact remnants produced in 
the implosion.  Section~\ref{sec:explosion} reviews the current explosion 
mechanisms and section~\ref{sec:bhformation} reviews how these explosions 
dictate the remnant mass distribution.  These sections cover theoretical 
expectations.  We conclude with a discussion of the current observations 
including the expectations from gravitational wave detections.

\section{Explosions from Stellar Collapse}
\label{sec:explosion}

\subsection{Convection-Enhanced Supernovae}
\label{sec:snengine}

As the iron core collapses, the subsequent higher densities and
temperatures drive further dissociation of the iron in the core
(removing thermal support) and increased electron capture (removing
electron degeneracy support) causing the collapse to accelerate.
Ultimately, the core collapses at over 1/10th the speed of light.
This nearly free-fall collapse continues until the core reaches
nuclear densities, where nuclear forces and neutron degeneracy halt
the collapse, producing a proto-neutron star (PNS)\footnote{As we shall
  discuss in Section~\ref{sec:vms}, the high entropy cores of very massive stars,
  the collaps halts before reaching nuclear densities and the core can
  collapse to a BH prior to reaching nuclear densities.}.  At
this time, a roughly 0.8-1.0\,M$_\odot$ iron core has collapsed from $\sim
5000$\,km down to $50$\,km.  The sudden halt of the collapse causes the 
infalling material to ``bounce''.

In the collapse, roughly $G M^2_{\rm PNS}/R_{\rm PNS} \approx
10^{53}$\,erg\footnote{In this equation, $G$ is the gravitational
  constant, $M_{\rm PNS}$ is the mass of the core (1-1.4\,M$_\odot$)
  and $R_{\rm PNS}$ is the radius of the collapsed core (typically
  this radius is initially 50\,km but as the PNS cools, it contracts to
  10\,km).} of gravitational potential energy is released (far above
what is needed to power typical supernovae).  Unfortunately,
converting just 1\% of this energy into explosion energy to make a
$10^{51}$\,erg supernovae\footnote{Hans Bethe termed this unit a
  ``foe'' for {\bf f}ifty {\bf o}ne {\bf e}rg, but it has also been
  termed a ``Bethe''.} has proven extremely difficult.  The first
simulations argued that the ``bounce'' would convert the potential
energy released into outward-moving kinetic energy, driving an
explosion~\cite{colgate66}.  However, much of the energy in this
bounce is stored in thermal/neutrino energy and when neutrino
transport is included, the bounce explosion stalls as soon as the
neutrinos are no longer trapped in the shock (roughly at 100-200\,km).

Over the past 40-50 years, astrophysicists have focused on studying
how to revive this shock.  The rest of the star falls onto the stalled
shock and the ram pressure of this infalling material must be overcome
to drive a successful explosion.  Both a region within the
PNS and the region between the PNS and
the stalled shock are susceptible to a number of convective
instabilities\footnote{Here, we define convection as any instability
  driving mass motions within a region, similar to the terminology
  used in stellar evolution.}.  These instabilities can increase the
efficency at which the potential energy is converted into explosion
energy to revive the shock and blow off the infalling star.  Much 
of the current work is focusing on the interplay of the enhancement 
in energy conversion from convection coupled to the increasingly complex physics 
needed to model core collapse (behavior of matter at extreme 
densities, neutrino emission and cross-section, neutrino transport, 
and magnetic fields).

\begin{figure}
\includegraphics[scale=0.5]{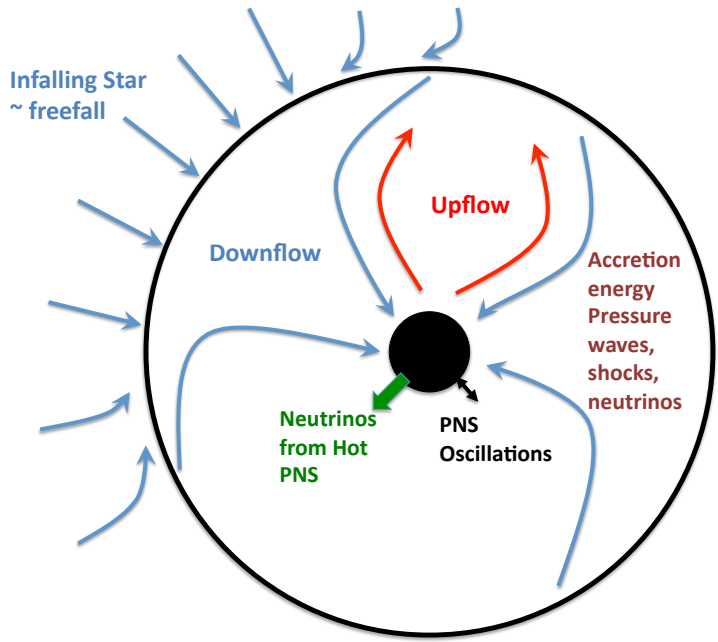}
\caption{Diagram of the supernova engine.  The collapsed core forms a PNS 
roughly 50\,km in size.  The bounce shock moves outward to 100-200\,km 
before stalling.  At this point, the rest of the star has realized its 
core has collapsed and falls onto this stalled shock.  That material 
builds on the stalled shock, flowing to low points.  After enough matter 
is built up, it sends downflows toward the PNS.  Material heated by 
a range of energy sources (neutrinos leaking out of the PNS or from 
accreting material, shocks on the PNS surface, PNS oscillations) rises 
in bubbles, converting the heat energy into kinetic energy that pushes 
out on the stalled shock.  If this energy can overcome the ram pressure 
of the infalling star, the stalled shock can move outward and drive an 
explosion.}
\label{fig:snconvdiag}
\end{figure}

Most of the current work is focused on the convection between the PNS
and the stalled shock~\cite{herant94}, where turbulence is able to
extract energy released near the surface of the NS and convert it to
kinetic energy driving out the shock (see
Figure~\ref{fig:snconvdiag}).  The energy in the convective region
must overcome the ram pressure of the infalling star.  This ram
pressure ($P_{\rm shock}$) of the infalling material rapidly decreases
with radius ($P_{\rm shock} \propto \rho v^2 \propto
r^{-3}-r^{-4}$\cite{fryer96}), so if the stalled shock can be forced
outward, it will quickly accelerate, launching an explosion.  Our
current focus is on understanding the nature of the turbulence
(e.g. whether it is driven by Rayleigh Taylor or standing accretion
shock instabilities), the source of the energy (hot proto-neutron
star, PNS oscillations, pressure waves from material accreting onto
the PNS) and what critical physics is missing in the current models
(e.g. neutrino interactions and transport, equation of state, magetic
fields).  All these topics remain areas of intense debate.  Fueled by
the fact that GRBs and hypernovae are almost certainly driven by a
different mechanism, many astronomers are studying mechanisms beyond
this convection-enhanced engine.

\begin{figure}
\includegraphics[scale=0.5]{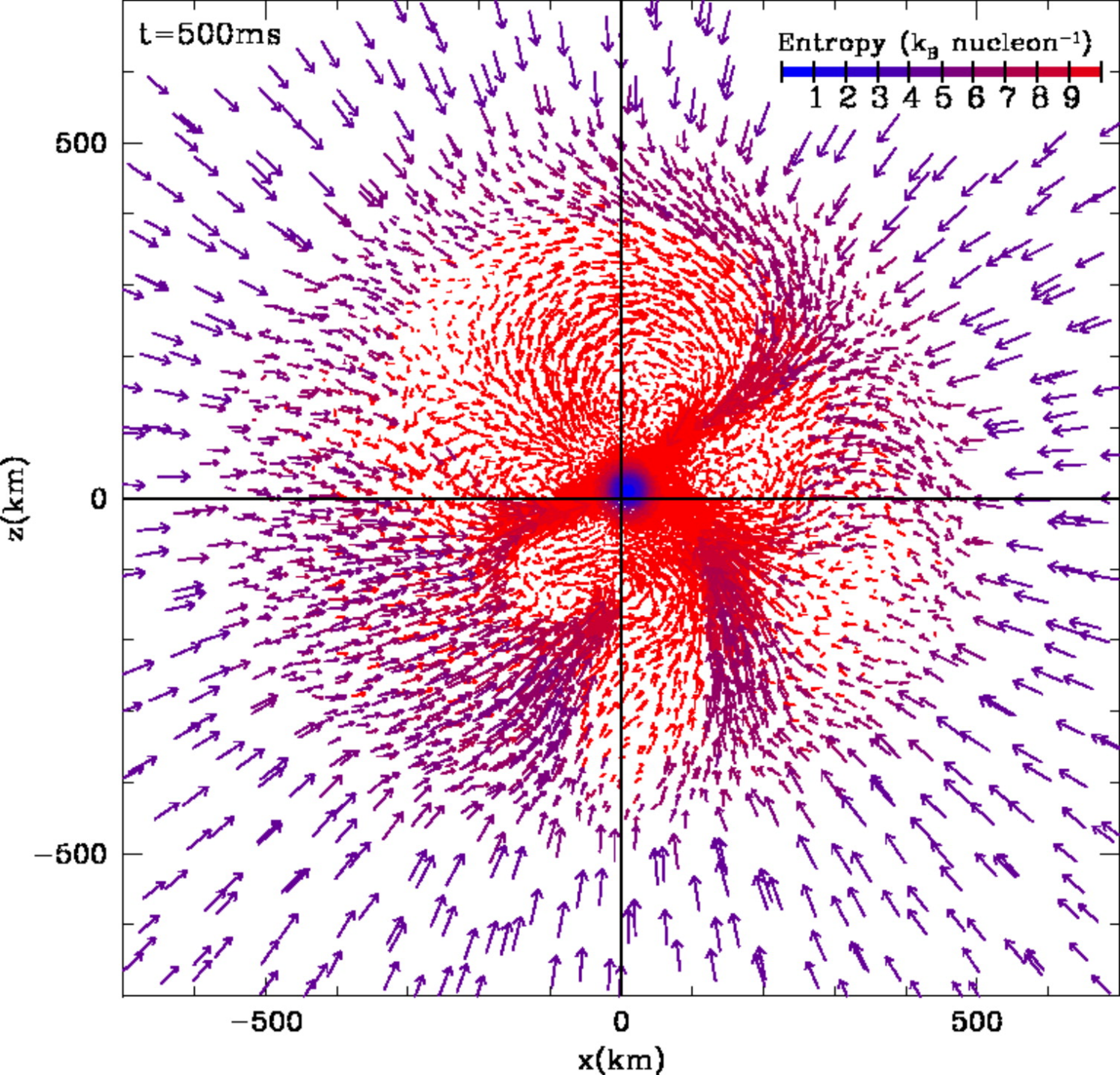}
\caption{Slice in the x-z plane of a 3-dimensional core-collapse
  simulation\cite{fryer07}.  The shading denotes entropy and the
  arrows denote velocity direction and magnitude.  As the star
  implodes, the outer layers of the star rain down on the collapsed
  core and flow around the upflows to form downflows.  The downflows
  crash onto the PNS, adding to the neutrino emission.  Shock energy
  also can increase the entropy of the material, contributing to
  upflows.}
\label{fig:snsim}
\end{figure}

Before we discuss these alternate mechanisms, let's study in more
detail some of the scenarios within the convection-enhanced engine
paradigm using figure~\ref{fig:snconvdiag} as a guide.  After the
collapse and bounce, unless there are significant pathologies in the
equation of state, the entropy profile from the PNS out to the shock
will decrease ($\partial S/\partial r<0$\cite{houck91,fryer96} and
such a profile is unstable to Rayleigh-Taylor
instabilities\cite{bethe90,herant94}.  The growth time of such
instabilities is roughly a few milliseconds\cite{fryer07} and we
expect Rayleigh-Taylor to dominate at early times.  Much of the
potential energy released in the collapse of the core is converted
into heat in the PNS and this heat leaks of of this star via
neutrinos.  These neutrinos heat material just above the PNS, further
increasing the entropy gradient to drive further Rayleigh-Taylor
convection.  The neutrino-deposited energy is converted into kinetic
energy as the hot bubbles rise from the PNS base, pushing against the
infalling star.  This convective engine is depicted in a slice of a
3-dimensional calculation in figure~\ref{fig:snsim}.  As the
collapsing star piles up on the stalled shock, it flows toward low
points, piling up until a downflow is produced.  The low entropy
material flows down onto the PNS, producing shocks and enhancing the
neutrino emission (contributing additional sources of energy).  If the
energy is sufficient to drive the shocked region outward, the ram
pressure will quickly decrease, allowing a strong supernova explosion.
As the explosion becomes strong, the convective engine turns off.  In
this way, the amount of energy in the explosion can be estimated by
setting it equal to the energy stored in the convective region when
the ram-pressure lid is blown off, allowing us to calculate fallback
and final remnant masses (Sec. ~\ref{sec:bhformation}).

Rayleigh-Taylor convection should occur in all simulations of stellar
collapse.  Indeed, calculations that don't exhibit Rayleigh-Taylor
convection typically either have some pathology in their equation of
state preventing a large initial entropy gradient, their resolution is
too low and numerical viscosity prevents the growth of the
Rayleigh-Taylor instability, or their initial conditions are
post-bounce and do not include an entropy gradient.  All calculations
suffer from low-resolution, explaining why, except in a few cases, the
Rayleigh-Taylor growth in the simulations is typically ten times
longer than the analytic predictions.  If the Rayleigh-Taylor
convection can quickly reset the entropy profile, other instabilities
can develop\cite{houck92}\footnote{Note, however that simulations of
  this profile show that such a resetting of the entropy profile takes
  longer than the 10 or so turnover times expected in rapid accretion
  onto NSs\cite{fryer07}.}.  It is this assumption that forms the basis
for the standing accretion shock
instability\cite{blondin03,blondin06}.  In this scenario,
entropy/vorticity perturbations are advected downward in the flow and
couple with acoustic perturbations.  This instability can drive
low-mode oscillations, again converting potential energy released near
the base of the PNS into kinetic energy at the edge of the stalled
shock.  Like the Rayleigh-Taylor instability, the energy in this
explosion can be estimated by the maximum energy stored in the
convective region.

All in all, scientists have focused on a few variants in instabilities:
\begin{itemize}
\item{Rayleigh-Taylor Instability: driven by entropy gradients,
  high-entropy (or ``hot'') bubbles will rise from the PNS surface.
  Low entropy material will find low points in the stalled shock, pile
  up, and form downstreams\cite{miller93,colgate93,herant94}.}
\item{Rotationally stabilized Rayleigh-Taylor, also known as the
  Solberg-H\o iland instability\cite{endal78}: if the specific angular
  momentum increases as one moves outward in radius (typical for
  rotating massive star structures), the angular momentum profile will
  stabilize against convection.  This can produce asymmetric
  explosions\cite{fryer00}.}
\item{Composition gradients and Rayleigh-Taylor: Composition
  gradients, e.g. gradients in the electron fraction in the
  PNS core can drive or stabilize against convection
  (if it counters the entropy gradient, the relative diffusion of heat
  and composition can drive a doubly diffusive instability).  This
  instability is studied primarily in convection within the
  PNS\cite{keil96,dessart06}.}
\item{Acoustic/Advective Instability (the standing accretion shock
  instability is a variant of this instability): This instability,
  coupling vorticity and pressure waves, is slower to grow than
  Rayleigh-Taylor but may develop and dominate in some
  scenarios\cite{blondin03,blondin06,foglizzo07}.}
\end{itemize}
and a range of energy sources:
\begin{itemize}
\item{Thermal energy in the PNS: (gravitational
  potential energy released in initial collapse).  The energy is
  radiated from the NS in the form of neutrinos\cite{zwicky38}.}
\item{Gravitational energy released from infalling material striking
  the PNS.  Shocked material may drive some material  
to rise or neutrinos from the accreting material can transfer energy.}
\item{PNS oscillations:  The accretion, especially in acoustic/advective 
instabilities, can drive oscillations in the NS\cite{burrows06}, although 
see\cite{fryer07,yoshida07}.}
\item{Rotational Energy: If the star is rotating rapidly, a disk can
  form around the PNS.  This rotational energy can drive a dynamo that
  generates strong magnetic fields to drive an
  explosion\cite{akiyama03,wheeler05}.  The rotation period during the
  engine for even the fastest-spinning stars is typically too long
  (70-770\,ms) to wind a strong magnetic field during the initial engine,
  but as the PNS contracts, this engine becomes viable\cite{fryer00,dessart08}.}
\end{itemize}

\subsection{Late-Time Energy Sources}

As the convective region expands with the launch of the explosion, the
amount of mass in this region that can absorb neutrinos from the hot
PNS decreases, effectively shutting off this source of energy.  PNS
oscillations can deposit energy longer, but even this energy source
diminishes.  However, new energy sources develop after the collapse.
Both post-explosion fallback and rotational energy have been invoked
to drive explosions.

Supernova ``fallback'' is material initially ejected in the supernova
explosion that ultimately falls back onto the PNS.  Two major
scenarios have been proposed to explain this fallback focusing on a
1-dimensional picture of supernova explosions.  The first argued that
the ejecta decelerates as it plows through the imploding star.  It
produces an explosion, but loses enough energy to fall back onto the
PNS\cite{colgate71}.  Alternatively, a reverse shock is produced when
the supernova shock enters the hydrogen envelope (with its shallow
density gradient).  This reverse shock drives
fallback\cite{woosley89}.  The primary difference between these two
scenarios is that the reverse shock scenario only produces fallback
after a considerable delay (the shock must first reach the hydrogen
envelope and then the reverse shock must decelerate the outward moving
material causing it to fall back).  The ejecta deceleration
model\cite{colgate71} tends to produce considerable fallback in the
first 1-15\,s\cite{fryer09}.

\begin{figure}
\includegraphics[scale=2.0]{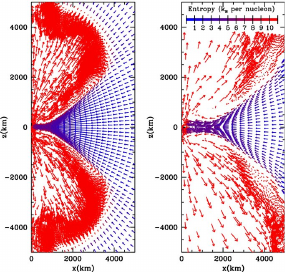}
\caption{An x-z image of a 2-dimensional simulation of outflows from fallback.  
The vectors dictate the direction and magnitude of the velocity and the shading 
denotes the entropy.  Some of the material receives enough entropy to rise and 
escape the downflow.  The initial accretion rate is $0.01 M_\odot {\rm \, s^{-1}}$ 
and there is a modest amount of angular momentum ($3\times10^{15} {\rm cm^2 \, s^{-1}}$) 
that provides a preferred ejection direction (along the rotation axis).}
\label{fig:fallback}
\end{figure}

The amount and nature of then fallback in 1-dimensional simulations
depends not just on the characteristics of the explosion (progenitor
star and explosion energy), but also on the numerical implementation
of the
explosion~\cite{macfadyen01,young07,zhang08,fryer09,ugliano12,dexter13}.
Indeed, those models using piston-driven explosions are limited to the
late-time reverse-shock fallback.  Energy injected models have more
fallback primarily in the first 1-15\,s after the launch of the
explosion~\cite{fryer09} (for a comparison of these two engines,
see~\cite{young07}).  In multi-dimensional simulations, fallback
occurs even as the shock is launched as low-entropy material flows
down through the hot rising shock.  This is seen both in late-time
explosion calculations and ejecta supernova explosions, making it
difficult to distinguish fallback from the supernova mechanism
itself~\cite{bruenn10,ellinger13}.  For this paper, we will call any
material accreting onto the NS after the launch of the shock
``fallback''.  This fallback shocks when it hits the cooling PNS and
some of the material is re-ejected in plumes
(Fig.~\ref{fig:fallback})~\cite{fryer96,fryer09}.  These plumes can
make up a sizable fraction of the explosion energy, especially in
those weak explosions that produce considerable fallback.  The amount
of fallback ties the nature of the explosion engine directly to the
remnant mass.  We will discuss fallback in much greater detail in
section~\ref{sec:bhformation}.

Rotating stars produce asymmetric explosions that tend to leave behind
disks around the PNS.  Magnetic fields are capable of extracting this
rotation energy to drive a second explosion.  After the launch of the
explosion, the PNS contracts.  Any rotation disk will contract,
speeding its rotation period and within a few seconds a dynamo can
create a strong magnetic field.  This magnetic field may produce
jet-like explosions with a NS compact
remnant\cite{dessart08}.

The inner core of a massive star often couples to outer layers,
causing its angular velocity to be on par with these outer layers.
This means the core's angular momentum is quite low.  If, however, the
PNS collapses to a BH and the inner $3-5 M_\odot$ accretes
onto the compact remnant, the higher angular momentum material of the
outer core can form a rapidly spinning accretion disk around the black
hole.  It is this disk that has been invoked to form the collapsar GRB
engine~\cite{woosley93,popham99}.  The collapsar model argues that a
magnetic dynamo in the disk (or neutrinos in the disk) produce a
highly-relativistic jet, producing a gamma-ray burst.  This jet
eventually disrupts the star, halting accretion and limiting the black
hole mass.

\subsection{Very Massive Stars}
\label{sec:vms}

As the mass of the star increases, a new instability can occur wherby
pair-creation in the core causes the star to collapse and undergo
explosive oxygen and silicon
burning\cite{barkat67,woosley82,bond84,carr84}.  For stars with masses
above $\sim 260 M_\odot$, a sufficiently large fraction of the center
of the star becomes so hot that photodisentragion occurs before
explosive burning reverses the implosion, accelerating the
collapse\cite{bond84}.  These high entropy cores halt the infall
before reaching nuclear densities, forming a proto-BH and
producing a very different neutrino signal from lower mass stars and
produce, at best, a disk wind outburst much dimmer than any
supernova\cite{FWH01}.  The proto-BH keeps accreting until an
event horizon forms around it.  Such proto-BH persists out to
masses above $100,000 M_\odot$\cite{FH11} and the fate can only be
altered if the core is rapidly rotating.

\section{Black Hole Formation and Masses}
\label{sec:bhformation}

There are 3 basic ways a BH forms in stellar collapse.  
\begin{itemize}
\item{Direct Collapse: In systems where all the explosion mechanisms
  discuss in section~\ref{sec:snengine} fail to revive the shock, the
  PNS continues to accrete and ultimately becomes too massive to
  support itself, collapsing to form a BH.  In these systems,
  the mass of the BH is the mass of the star at collapse.
  Fryer\cite{fryer99,heger03} argued that stars above $\sim 40 M_\odot$
  (neglecting mass loss from winds and binary interactions) collapse
  directly to BHs.  More recent estimates predict a lower
  limit but with large variations caused by stellar mass loss\cite{fryer99}.}
\item{Fallback Collapse:  In systems where a weak explosion occurs, much 
of the material initially ejected at the launch of the explosion falls back, 
causing the PNS to collapse to a BH~\cite{fryer99}.  This 
formation scenario can form a wide range of masses for low-mass BHs.}
\item{Proto-BH Formation: Very massive stars (above $\sim 300
  M_\odot$) produce a proto-BH that accretes mass until an
  event horizon forms, causing it to become a true black
  hole\cite{FWH01}.  Like the direct collapse stars, these stars form
  BH masses set to the stellar mass at collapse. }
\end{itemize}

\begin{figure}
\includegraphics[scale=0.5]{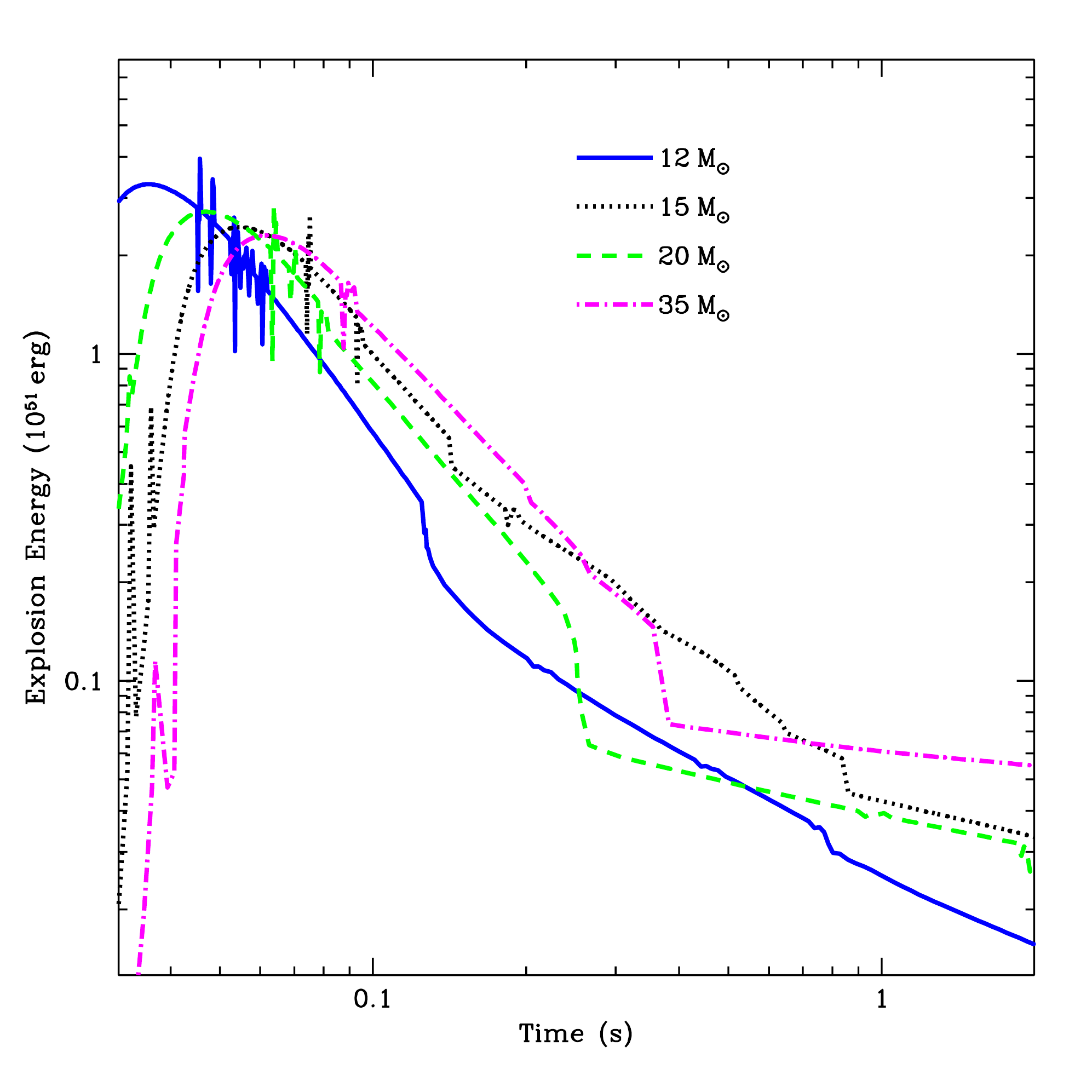}
\caption{Supernova explosion energy as a function of the time it
  takes, post-bounce, for the shock to be revived~\cite{fryer12}.  The
  density drops rapidly in the star as one moves from the silicon core
  through the carbon/oxygen and helium layers.  This drop in density
  converts to a drop in mass infall rate ($\dot{M}$) when the star
  implodes.  The total energy stored is proportional to the ram
  pressure ($P_{\rm ram}$) that, in turn, is set by the mass infall
  rate: $P_{\rm ram} = 1/2 \rho_{\rm infall} v_{\rm infall}^2 =
  (\dot{M} v_{\rm infall})/(8 \pi r_{\rm shock}^2)$ where $\rho_{\rm
    infall}$ and $v_{\rm infall}$ are, respectively, the infall
  density and velocity and $r_{\rm shock}$ is the radius of the
  stalled shock.  As the carbon/oxygen shell becomes the dominant
  infalling material, the pressure drops considerably.  If the
  convective engine has stored up $10^{51} {\rm erg}$ of energy in the
  turbulent region, it will explode.  After this time, the maximum
  energy that can be stored in the convective region is well below the
  typical $1-3 {\rm foe/Bethe}$ supernova energy (see\cite{fryer12}
  for more details).}
\label{fig:snenergy}
\end{figure}

Most ``normal'' supernovae have explosion energies between
$1-3\times10^{51}\,{\rm erg}$, only 1\% of the total potential energy
released in the collapse.  The convective engine provides a natural
explanation for this 1\% efficiency: because energy deposition drops
as soon as the shock is launched, the explosion energy is set to the
energy stored in the convective region prior to the launch of the
shock (see\cite{fryer12} for more details).  In this manner, the
explosion energy depends upon the ram pressure from the infalling
stellar material; a function of both the stellar progenitor and the
time of explosion.  Figure~\ref{fig:snenergy} shows the explosion
energies as a function of explosion times for 3 progenitor masses
(assuming no mass loss).  The maximum energy stored in the convective
engine is never more than a few times $10^{51} {\rm erg}$, explaining
why the energy in normal supernovae are only 1\% the total energy
released\footnote{Hypernove and GRBs can extract up to 10\% the total
  energy released, clearly requring a different/additional energy
  source beyond the convective engine.}.  These strong explosions have
very little fallback (less than a few tenths of a solar mass) and the
remnants are typically NSs.

\begin{figure}
\includegraphics[scale=0.4]{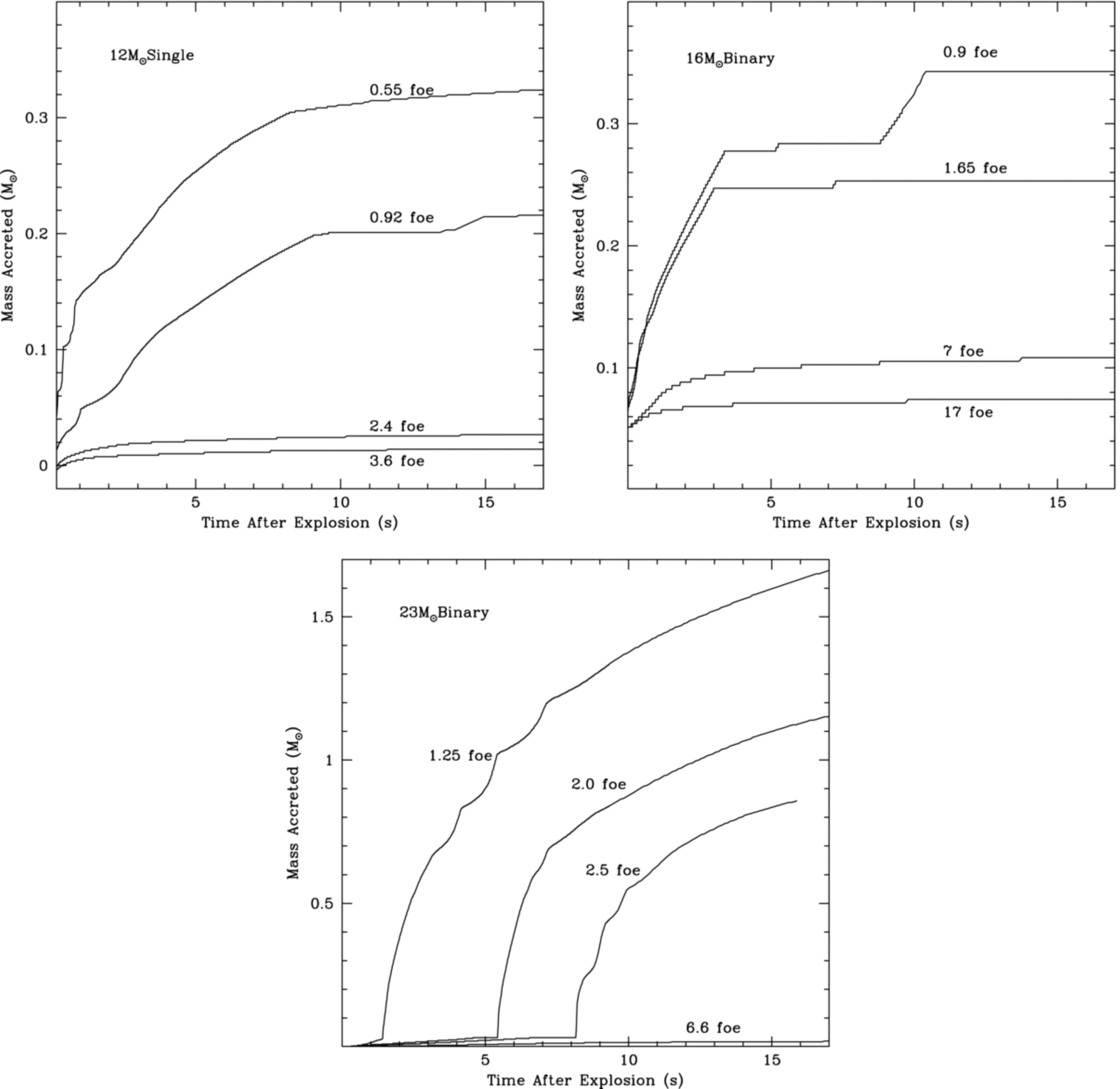}
\caption{Amount of fallback for 3 different progenitors: $12 M_\odot$
  single star and $16 M_\odot$ and $23 M_\odot$ binary stars (in the
  binary systems, the hydrogen is removed during a common envelope
  evolution phase, removing reducing the mass)\cite{fryer09}.  Note
  that the fallback begins within the first second and most occurs
  within the first 15\,s.  For stars with masses above $20 M_\odot$,
  there is considerable fallback (more than $0.5 M_\odot$) for any
  explosion with normal (1-3foe) explosion energies.}
\label{fig:fallbackmass}
\end{figure}

Although this naturally explains most supernovae, if the explosion is
delayed (taking more than $100 {\rm ms}$ after bounce), explosion
energies are typically less than $10^{51} {\rm erg}$.  These weaker
explosions lead to considerable fallback, producing a range of remnant
masses from massive NSs ($2 M_\odot$) to BHs with masses
nearly as massive as those produced through ``direct'' collapse.
Supernovae with considerable fallback provide a unique probe into the
supernova engine.  Figure~\ref{fig:fallbackmass} shows the mass of the
fallback material as a function of time for multiple explosion
energies and 3 different stellar progenitor masses.  Most of the
fallback happens in the first 1-15\,s after the launch of the
shock\footnote{This means that any additional energy from fallback
  will be able to catch up to the shock before the shock breaks out of
  the star, contributing to the observed supernova explosion energy.
  This may be alternate explanation for bright supernovae with low
  $^{56}$Ni, e.g. SN 2005bf\cite{maeda05}.}.  Above $\sim 20 M_\odot$,
the explosion energy must be very strong to avoid significant
fallback.  For example, a $2\times 10^{51}{\rm erg}$ (2 foe) explosion
for our binary $23 M_\odot$ progenitor has over $1.1 M_\odot$ of
fallback in the first 15s.

\begin{figure}
\includegraphics[scale=0.4]{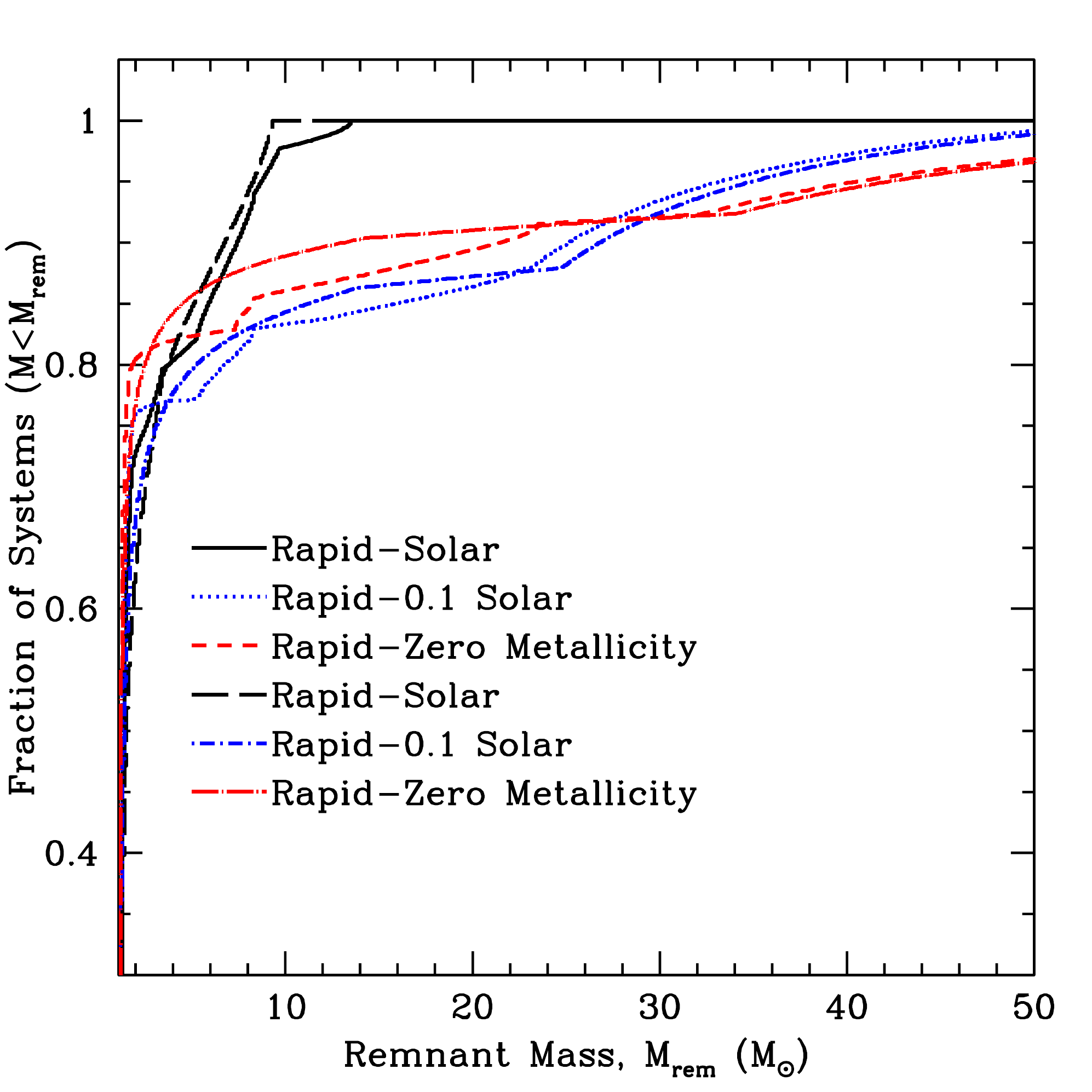}
\includegraphics[scale=0.4]{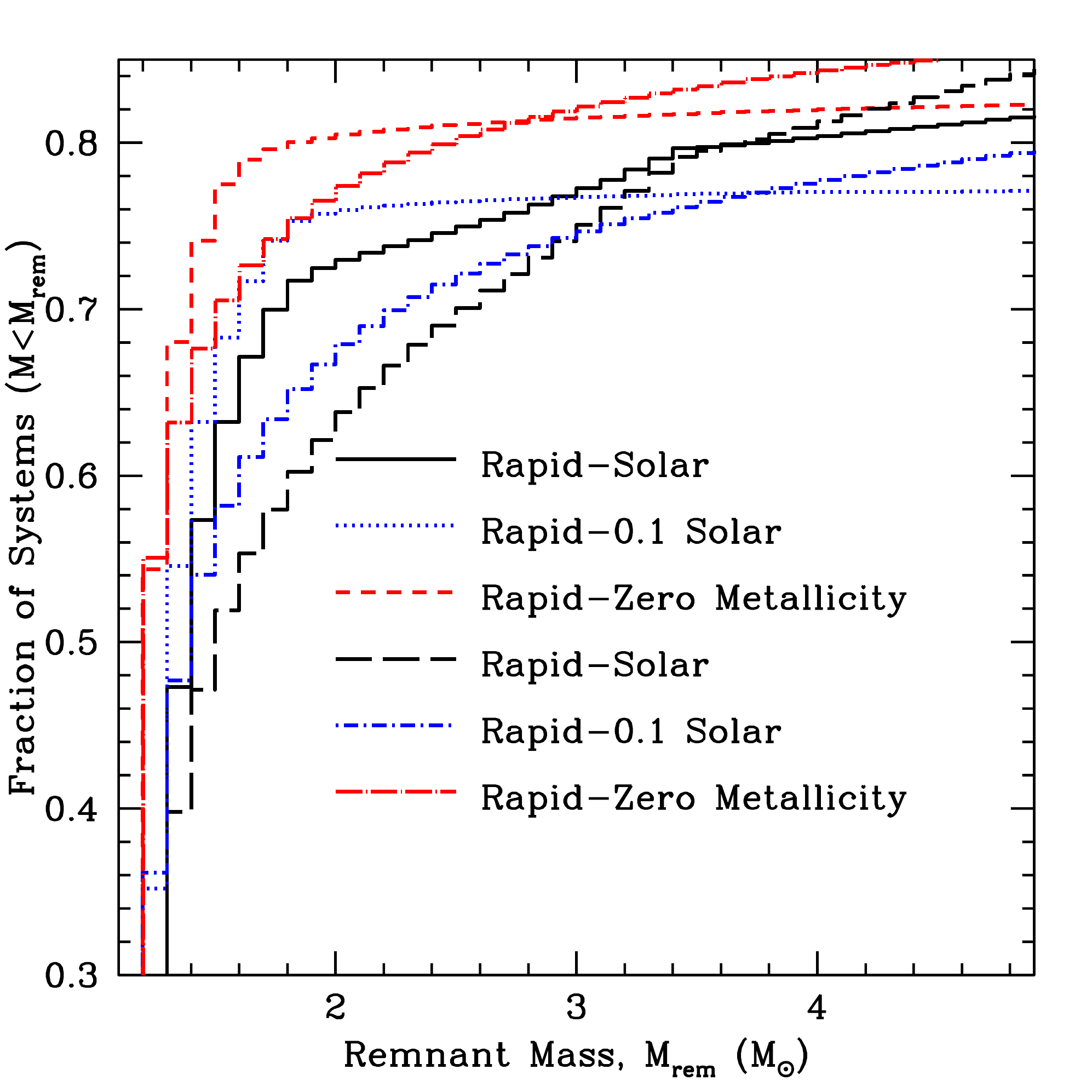}
\caption{Cumulative fraction of remnant masses as a function of mass
  for 6 models: 3 rapid and 3 delayed explosions each at solar, 0.1
  solar and zero Metallicity.  In these models, 80-90\% of all compact
  remnants are NSs.  For the rapid explosions, 75-95\% of these NSs
  have masses below $1.6 M_\odot$.  Very few remnants are produced
  with masses between $2 M_\odot$ and $4 M_\odot$.  For the delayed
  explosion, a more distributed range of masses exist.}
\label{fig:nsmass}
\end{figure}

The cumulative distribution of remnant masses for both delayed and
rapid engine models for 3 different metallicities is shown in
figure~\ref{fig:nsmass}.  In these models, 80-90\% of all compact
remnants are NSs.  For the rapid explosions, 75-95\% of these NSs have
masses below $1.6 M_\odot$.  Very few remnants are produced with
masses between $2 M_\odot$ and $4 M_\odot$.  For the delayed
explosion, a more distributed range of masses exist.  With these
differences, accurate observations of the mass distribution will
determine whether a rapid or delayed explosion dominates the supernova
engine (shedding light into the convective instability) and whether 
a new energy source is needed beyond that available in the convective 
region.

\subsection{Compact Remnant Kicks}

Most observations of compact remnants are focused on compact remnants
in binaries.  Binary observations are a biased sample, limited to
those systems that remain bound after the formation of the compact
binary.  The violent collapse of a stellar core can impart large
velocities on the compact remnants.  To understand these systems, we
must understand the ``kicks'' imparted on these remnants at birth.

A growing set of both NS and BH binaries seem to
require that the compact remnant receive a burst of momentum at
formation\cite{fryerkalogera98,fryer98,gualandris05,willems05,fragos09}.
Observations of pulsar proper motions suggest that the distribution of
these kicks contains a sizable fraction of neutrons stars moving with
velocities in excess of $400 {\rm km \, s^{-1}}$\cite{cordes98,arzoumanian02}.  Although
uncertainties persist in these measurements (distance errors, issues
such as biases in the sampling - systems with stronger kicks are more
likely to escape the galaxy), these observations place fairly strong
constraints on kicks received by NSs.  Much less is known
about the BH kick distributions; aside from a few binaries
that suggest kicks are needed for some systems, we have very little
information on BH kicks\cite{willems05,fragos09}.

Although there exist many theories to explain kicks, nearly all have
trouble explaining those pulsars with velocities in excess of $1000
{\rm km \, s^{-1}}$ and although there is a preference among theorists
towards kicks produced by asymmetries in the supernova ejecta, no kick
mechanism can truly be ruled out or in.  Most of the kick mechanisms
invoke asymmetries in the supernova explosion whereby momentum
conservation requires that if the ejecta is driven preferentially in
one direction, the compact remnant must be directed in the opposite
direction with equal momentum.  This means that both the mass and
velocity of the ejecta asymmetry is critical.  In this way, GRB jets
have difficulties driving kicks.  With only $\sim 10^{-6} M_\odot$ in
the jet mass, even if the jet is one-sided with a lorentz factor of
500, the kick on a $5 M_\odot$ BH is only $30 {\rm km
  \, s^{-1}}$.  But a number of both ejecta and neutrino asymmetries
could produce large kicks:
\begin{itemize}
\item{Binary disruption: The sudden ejection of the supernova can
  disrupt a binary star system.  If the supernova has a lot of ejecta
  and the binary is close, this scenario produces kicks up to $200
  {\rm km \, s^{-1}}$, well below the bulk of the pulsar
  velocities\cite{gott70}.  The ejecta in a binary can also impart
  bulk velocities on the bound binaries, but these velocities are
  extremely low and can not explain the fastest binary systems\cite{willems05,fragos09}.}
\item{Propeller: Asymmetric emission of radio waves for pulsars where
  the magnetic field is off-centered and inclined to the axis of
  rotation\cite{harrison75}.}
\item{Growth of Pre-Collapse Asymmetries: Although it has been argued
  that asymmetries in the initial structure could drive strong kicks,
  realistic calculations show that only abnormally large asymmetries
  can produce kicks above a few hundred ${\rm km
    \, s^{-1}}$\cite{fryer04}.}
\item{Low-Mode Rayleigh Taylor: As the Rayleigh-Taylor instability
  develops, the lowest modes grow roughly to the size of the stalled
  shock (Fig.~\ref{fig:snsim}).  Such low modes can drive asymmetries
  in the supernova explosion ejecta, producing kicks\cite{herant95}.  This
  mechanism has received some success\cite{buras03}.}
\item{Advective/Acoustic Instabilities:  Low-mode instabilities are 
characteristic in this instability and several groups have studied the 
kicks from these instabilities\cite{}.}
\item{Neutrino Asymmetry from Neutrinosphere Turbulence:  Magnetoacoustic 
instabilities producing neutrino bubbles akin to the photon bubble instability 
produces asymmetric neutrino emission as the neutrino bubbles rise to 
the neutrinosphere\cite{socrates05}.}
\item{Neutrino Oscillations: If the core has extremely strong magnetic
  fields, neutrino oscillations coupled to those magnetic fields, can
  produce strong neutrino
  asymmetries\cite{kusenko96,kusenko97,barkovitch02,barkovitch04,fuller03,kusenko04,fryerkusenko05}.}
\end{itemize}  

Unfortunately, none of these kick mechanisms are well-enough understood 
to then infer kick distributions for BHs.  But the kick placed on 
the BH can be categorized into 3 fates\cite{whalen12}:
\begin{itemize}
\item{No Kick:  the propeller mechanism for all BHs, binary disruption 
and all ejecta explosions for direct BHs, neutrino-driven kicks may work 
in all BH formation scenarios.}
\item{Equal Momentum Kicks: Binary and ejecta-driven kick mechanisms
  are likely to produce kicks with equal momenta to their NS
  counterparts, hence the kick velocity is roughly $\propto M_{\rm
    NS}/M_{\rm BH} \times v_{\rm NS kick}$ where $M_{\rm NS}$, $M_{\rm
    BH}$ are the NS, BH masses respectively and
  $v_{\rm NS kick}$ is the NS velocity (taken from the
  proper motion studies of pulsars.  For neutrino driven mechanisms,
  this case may hold for both fallback and direct collapse BH
  scenarios.}
\item{Enhanced Momentum Kicks:  For convective-instability kick mechanisms, 
asymmetries are strongest in delayed explosions, exactly those explosions 
that produce a lot of fallback.  These systems might get the strongest kicks.}
\end{itemize}
For BH kicks, a few scenarios are typically implemented into
population studies of binary BH systems: BHs receive no kicks
whatsoever (there is a growing belief that this is disproved by
observations - e.g.\cite{willems05,fragos09}), all BHs receive kicks with
conserved momentum: $v_{\rm kick}\propto M_{\rm NS}/M_{\rm BH} \times
v_{\rm NS kick}$, direct BHs receive no kicks and BHs
with fallback receive kicks (either with conserved momentum, or in the
enhanced case, with the same velocity as the pulsar observations).  
If one of these can be demonstrated absolutely, we would have constraints 
on the kick mechanism for both NSs and BHs.  In addition, 
these kick distributions must be understood to tie theoretical mass distributions 
to observations of binary systems.

\section{Observational Constraints:  Supernovae and  Compact Remnant Masses}

Currently, the primary constraints on the mass distributions of
compact remnants is limited to binaries containing NSs and BHs
(e.g. X-ray binaries and binary pulsars).  Early analyses of these
binaries suggested that NS masses clustered with a very narrow mass
range around $1.35 \pm 0.04 M_\odot$\cite{thorsett99} and BHs
clustered with a broader range ($\sim 1 M_\odot$) around $7
M_\odot$\cite{bailyn98}.  But as new systems have been added and the
data on existing systems has become more refined, it has become clear
that the distribution of NSs is more spread out with neutron
star masses ranging from $\sim 1.1 M_\odot$\cite{vankerkwijk95} to
$\sim 2M_\odot$\cite{demorest10}.  BH mass measurements are even more
difficult to make, relying upon a complex combination of both
observations of X-ray binaries and modeling of photometric and
spectroscopic data.  Although the BH mass range has widened
with time, one aspect of the mass distribution continues to persist:
the existence of a gap between the NS masses (ending around
$2 M_\odot$) and BH masses (roughly $4 M_\odot$ and
above)\cite{ozel10,farr11}.  However, bear in mind that the gap is
statistical in nature and some BHs have error bars that fall
within this gap, but even if this gap is not pristine, it is likely
that there is a dearth of remnants with masses between $2-4\,M_\odot$.

\begin{figure}
\includegraphics[scale=0.5]{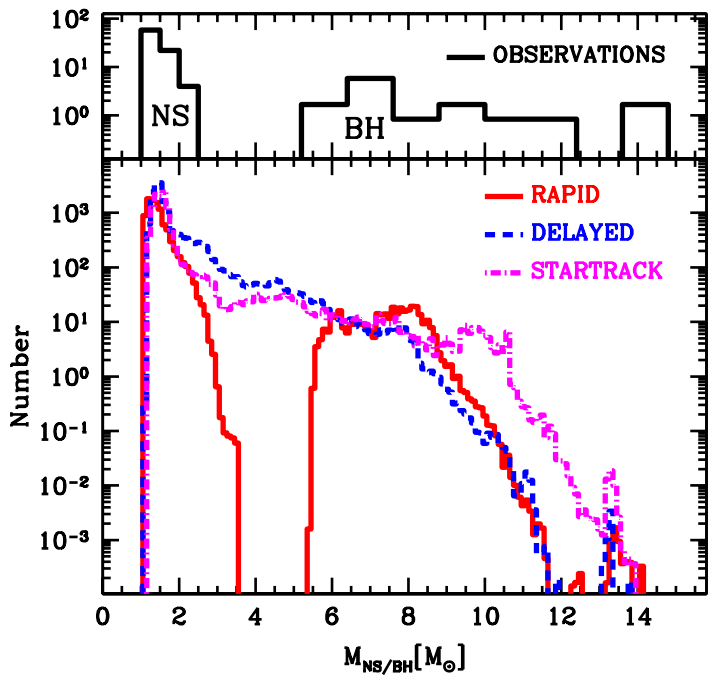}
\caption{Neutron star/black hole mass distribution of the Galactic population 
of Roche lobe overflow and wind fed X-ray binaries (top) with the predicted 
rates for 3 different models: rapid, delayed, and old formalism used in 
STARTRACK\cite{belczynski12}.  Note that only the rapid explosion model 
produces the observed gap.}
\label{fig:belczyn}
\end{figure}

The distribution of remnant masses, along with theory predictions
assuming rapid or delayed explosions, is shown in
figure~\ref{fig:belczyn}.  The top panel shows the observed
distribution (using the most-likely masses), clearly showing the
observed mass gap.  The bottom panel shows the number distribution for
3 different models: rapid explosion, delayed explosion, and former
model used by the STARTRACK code\cite{belczynski12}.  Only the rapid
explosion model produces a gap, suggesting that the turbulent engine 
must grow quickly.  This result supports a Rayleigh-Taylor instability.

\begin{figure}
\includegraphics[scale=0.5]{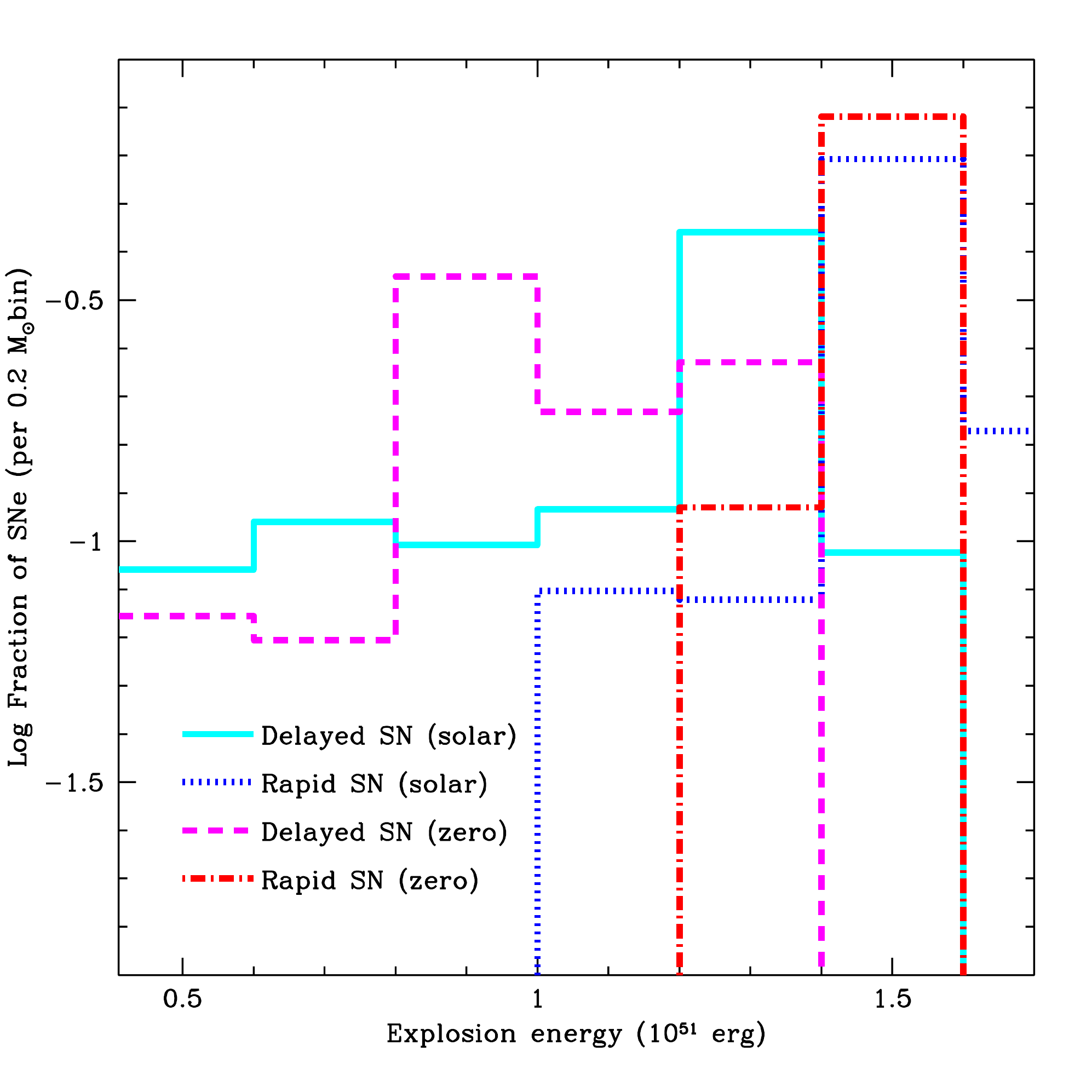}
\caption{Distribution of supernova explosion energy for our delayed
  and rapid explosions at two metallicities (solar and zero
  metallicity).  The rapid explosion mechanism produces only
  explosions in the 1-2foe energy range, whereas the delayed mechanism
  produces a wide range of supernova explosion energies.  The narrow
  range in the rapid explosion is why it can produce a gap: either it
  produces a strong explosion or it fails to make an explosion.
  Delayed engines produce much more fallback, filling in the gap.  }
\label{fig:sne}
\end{figure}

\begin{figure}
\includegraphics[scale=0.5]{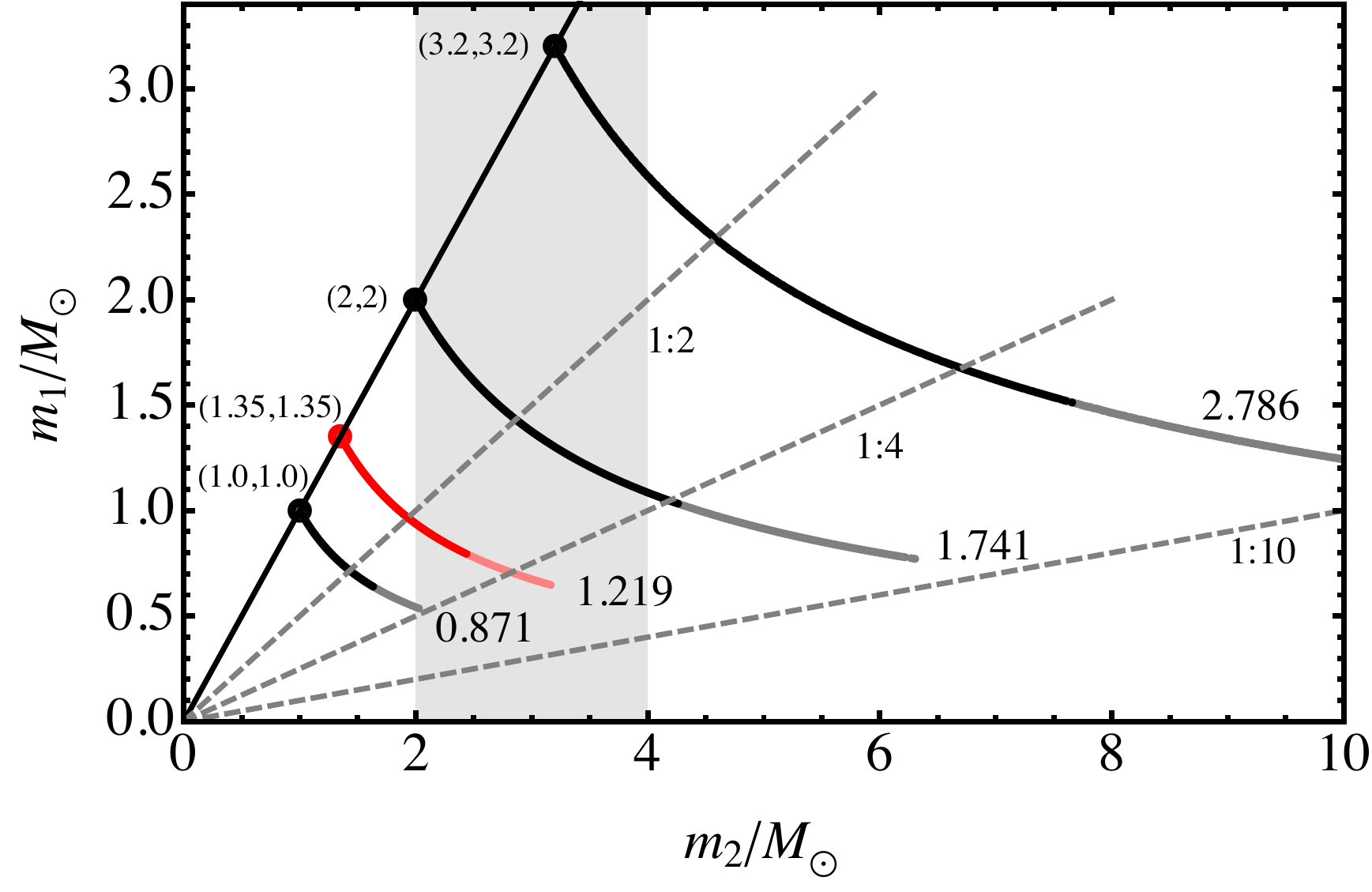}
\caption{ 90\% confidence regions for a number of different
  non-spinning compact-binary configurations. The number at the end of
  each confidence region is the chirp mass of the binary. The gray
  shaded region indicates the current observational mass
  gap\cite{brown13}.  Because gravitational waves do not measure the
  component masses, it is difficult to constrain component masses to
  better than 10-30\%.}
\label{fig:gw}
\end{figure}

But this is not the end of the story.  The current paradox with this
gap is that if it truly exists, it argues against many supernovae with
a large amount of fallback.  The supernova energies produced by our
two simplistic models at zero and solar metallicity are shown in
figure~\ref{fig:sne}.  The rapid explosion model which produces a gap,
does not produce any weak fallback supernovae.  But such supernovae
seem to exist\cite{moriya10}.  How can we have weak supernovae but no
fallback?  Barring a theoretical misunderstanding (possible), one of
the results, or more likely our interpretation of the results, is
incorrect.  Perhaps there are remnants with masses within the ``gap'' region, 
but they are sufficiently rare that they have yet to be observed (requiring 
a bigger sample).  This would allow some weak supernovae.

The primary difficulty in using the compact remnant masses to
constrain the supernova engine is the small sample statistics and
potentially biased data of the X-ray binary population.  Gravitational
waves, with the potential to double the number of systems in a few years, 
can provide new insight into the compact remnant distribution.  Unfortunately, 
although gravitational waves accurately measure the chirp mass, it is much 
more difficult to calculate the component star masses\cite{brown13}.  Figure~\ref{fig:gw} 
shows the range of possible inferred masses for a series of binary mass 
pairs (4 different chirp masses).  As with our X-ray binary systems, 
we will have to use large-sample statistics to demonstrate a ``gap'' in 
the compact remnant masses. Unfortunately, gravitational wave observations 
are limited to binary systems.  Although it is likely that gravitational 
waves will determine whether the gap truly exists, it will be difficult 
to determine whether this gap only exists in binary systems.

In X-ray binaries and binary compact mergers (gravitational wave
detections), the inferred BH mass is the mass of the BH at formation
plus the additional mass accreted in the binary.  Measurements of
solitary BHs will provide a more direct measurement of the BH mass
distribution at formation.  In addition, binary systems may be a
biased set (e.g., BHs that receive strong kicks, perhaps the lowest
mass BH systems, have a lower probability of remaining bound).  The
biases in BH binary formation are not well understood.  A Hubble Space
Telescope program is ongoing to look for single BHs using
microlensing.  WFIRST has the potential to do these studies and
increase our understanding of single BH masses.  

The connection between supernovae and compact remnants places 
us in an ideal position to use observations of one to constrain 
the other.  The next decade should bring about many new discoveries 
in these fields.

\end{document}